# Finite-Element Simulation of Casimir Forces in Arbitrary Geometries

M. Tajmar*

Space Propulsion
ARC Seibersdorf research, A-2444 Seibersdorf, Austria

**Abstract**

A 3D finite-element numerical simulation was developed to investigate Casimir forces in arbitrary geometries. The code was verified comparing it with results obtained from analytical equations. Appling the simulation to previously not assessed configurations, new Casimir properties were found such as repulsive Casimir forces in groove like structures.

* Principal Scientist, Email: martin.tajmar@arcs.ac.at, Tel: +43-50550-3142, Fax: +43-50550-3366

## Introduction

Vacuum forces acting on µm and sub-micron scales have been first calculated by Casimir and Polder[1,2] in 1948. One prediction was that two parallel plates will be actually attracted to each other due to the exclusion of electromagnetic modes between the plates compared to free space. Another interpretation is that the two plates lower the energy content of vacuum between them and that the vacuum outside the plates is pushing them together.

It took some 50 years until this remarkable effect of “vacuum engineering” was actually measured[3] for separation distances of 0.6 – 6 µm. With the introduction of nanotechnology, this effect becomes more and more important as force densities of as large as 120 kN/m$^2$ are predicted for separation distances of 10 nm. Recently, even a Microelectromechanical Systems (MEMS) structure was activated using Casimir forces[4].

Throughout the literature, almost all papers deal with two simple geometries: plate-plate and plate-sphere. Apart from first numerical approaches for more complex geometries[5-7], the first finite-element software calculating Casimir forces for planar structures from user-defined structures resulted from a contract by the European Space Agency to stimulate research on vacuum forces from an engineer’s perspective[8]. In addition to the planar-structure limit, the software was implemented in the interpreter language IDL, making it rather slow.

This paper presents the results of a fast 3D finite element code implemented in C for Casimir force calculations of arbitrary structures, allowing for more complex geometries such as wires and grooves. The software can calculate forces as well as the momentum of each object within the simulation domain. It also allows for simulating different materials for each object and their influence on the predicted force as well as to plot the Casimir potential distribution to illustrate the Casimir forces acting on the various structures. Results were compared with analytical equations and the software was applied to cases such as grooved structures which have not been previously assessed. Also for the first time, we use the code to assess the influence of the plate thickness in the classical plate-plate geometry.

## Numerical Approach

The code is based on a finite-element approach where a simulation domain is subdivided by a mesh structure. For simplicity, we assume a homogenous rectangular grid structure with a size in X, Y, and Z-direction. In our calculations, we assume that the object is continuous in Z-direction. Hence, the code calculates forces per length unit. First, the simulation domain is filled with objects of user defined size and material properties. The software supports the following objects shapes:

- Rectangular (Open, Filled),
- Triangular (Filled),
- Round (Open, Filled).

The Casimir forces on each object are computed using the Casimir-Polder interaction energy between two polarizable atoms A and B, which is given by[9]

$$U(r) = -\frac{\hbar c}{4\pi r^7}\left[23\cdot\left(\alpha_E^A\alpha_E^B + \alpha_M^A\alpha_M^B\right) - 7\cdot\left(\alpha_E^A\alpha_M^B + \alpha_E^A\alpha_M^B\right)\right] , \tag{1}$$

where $r$ is the distance between the particles and $\alpha_E$ and $\alpha_M$ are the electrostatic and magnetic polarizability of the atoms respectively. The magnetic polarizability can cause repulsive Casimir force[9], however, as for most materials the $\alpha_M$ is orders of magnitude below $\alpha_E$, we will approximate Equ. (1) by

$$U(r) \cong -\frac{23\hbar c\alpha_E^A\alpha_E^B}{4\pi r^7} . \tag{2}$$

The effect of magnetic polarizability can be implemented very easily at a later stage in the code by replacing Equ. (2) with Equ. (1). According to the object's materials property, one gridcell of object A will represent $N_A$ atoms (accordingly for the other objects). Therefore, the energy $U$ of an atom within a gridcell $i$ of object A can be expressed by summing up the interaction with all other gridcells $j$ from all other objects B, C, …

$$U_i = -\frac{23\hbar c\alpha_E^A N_A}{4\pi r^7}\cdot\left(\sum_j \frac{\alpha_E^B N_B}{\left|r_i - r_j\right|^7} + \sum_j \frac{\alpha_E^C N_B}{\left|r_i - r_j\right|^7} + \ldots\right) . \tag{3}$$

This additive approximation method is known to give the right dependency of $U$ with distance, however, the absolute value is usually not correct due to the non-additive effects connected with the screening of more distant layers of material by closer ones[10]. Therefore, a normalization procedure was suggested to approximately account for these non-additive effects[11]. According to the procedure, Equ. (3) should be divided by a normalization factor which is obtained by the division of the additive result by the exact quantity found from the plane-parallel configuration (analytical expression). We shall keep that in mind when comparing the solutions without normalization to the analytical expressions.

For plotting the effect of Casimir forces on the geometry, it is helpful to use the potential $\varphi_i$ in each gridpoint of the simulation domain, which is calculated by

$$\varphi_i = \frac{U_i}{\alpha_E^A N_A} \quad . \tag{4}$$

The force can be calculated from the energy by

$$\vec{F} = -\frac{d\vec{U}}{d\vec{x}} \quad , \tag{5}$$

or numerically the force per length unit *f* (in Z direction) can be expressed for its X and Y component by

$$f_x(X,Y) = \frac{U(X-1,Y) - U(X+1,Y)}{2 \cdot dx \cdot dz}, \; f_y(X,Y) = \frac{U(X,Y-1) - U(X,Y+1)}{2 \cdot dy \cdot dz} \quad , \tag{6}$$

where X and Y are the position on the grid. Only one Z layer is evaluated, the objects are assumed to expand both in the positive and negative direction with the size of the Z-simulation domain.

To remove numerical artefacts, for instance non-symmetric object gridpoints in the case of a spherical object, each object was calculated alone as a reference (each object alone should experience no force). This reference was then subtracted from the result of the object in the final configuration including all other objects.

## Results

To validate the code, simulation has been done on a plane-plate configuration which can be also expressed analytically. The force per area for perfectly conducting metal plates separated by the distance *r* is given by[3]

$$\frac{F}{A} = \frac{\pi^2}{240} \frac{\hbar c}{r^4} \quad . \tag{7}$$

For calculating real cases, two important correction factors have to be considered in Equ. (7): finite temperature and finite conductivity. The correction factor for finite temperature is given by[12]

$$f_c^T(T,r) = \left(1 + \frac{720}{\pi^2} f(\xi)\right) \quad , \tag{8}$$

where $\xi = kTr/\hbar c$, and

$$f(\xi) = \begin{cases} \left(\frac{\xi^3}{2\pi}\right) \cdot 1{,}202 - \left(\frac{\xi^4 \pi^2}{45}\right) & \xi \le 0.5 \\ \left(\frac{\xi}{8\pi}\right) \cdot 1{,}202 - \left(\frac{\pi^2}{720}\right) & \xi > 0.5 \end{cases} \quad . \tag{9}$$

The correction factor for finite conductivity (only applicable in case of metal surfaces) is given by[12]

$$f_c^C(\omega_p, r) = \left[1 - \frac{16}{3}\frac{c}{\omega_p r} + 24\left(\frac{c}{\omega_p r}\right)^2\right] . \tag{10}$$

Hence, the corrected analytical expression for plate-plate metal Casimir forces is given by:

$$\frac{F}{A} = \frac{\pi^2}{240}\frac{\hbar c}{r^4} \cdot f_c^T(T, r) \cdot f_c^C(\omega_p, r) . \tag{11}$$

For the numerical simulation, the force per length was calculated for two gold plates separated by 1 – 6 µm. The simulation parameters are listed in **Table 1**. Each run in this configuration takes about 350 s on a Pentium-II computer with 1.3 MHz. A comparison between the obtained forces (no normalization factor applied) with Equ. (7) and the corrected Equ. (11) is shown in **Fig. 1**. The numerical results show a similar shape compared to the analytical result in Equ. (7), however, the absolute value is underpredicted by about one order of magnitude. This shows clearly that the concept of re-normalization would indeed make sense and should be applied for obtaining accurate results.

**Fig. 2** shows the Casimir potential of Equ. (4) for the result in **Fig. 1** with a plate-separation of 2 µm. The changing potential between the plates makes the Casimir force visible. By increasing the simulation domain to 30x30x30 µm, it is clearly evident that Casimir forces are a very local phenomenon as the potential distribution becomes quickly homogenous (see **Fig. 3**). The influence of the force computation on the number of gridpoints was investigated in **Fig. 4**, varying the number of gridpoints from 20 to 200 on each axis (this validates our choice of 100 gridpoints used in the computations). Also the influence of the plate thickness on the results was investigated as shown in **Fig. 5**. The result can be actually fitted with an exponential decay function, leading to a force per unit length of $5.32 \times 10^{-11}$ $Nm^{-1}$ for infinite plate thickness compared to $3.87 \times 10^{-11}$ $Nm^{-1}$ for a plate thickness of only 1 µm (used for the computations in **Fig. 1**).

With the confidence of the results obtained above, new geometries were assessed that can not be calculated analytically – or at least only with great difficulty. First, a wire was put next to the bottom of a plate (both made out of gold) as shown in the Casimir potential distribution plot of **Fig. 6**. The forces acting on the wire are $f_x=1.55 \times 10^{-10}$ $Nm^{-1}$ and $f_y=1.2 \times 10^{-10}$ $Nm^{-1}$. The forces acting on the plate are exactly the same with opposite signs. So in addition to experiencing a push towards the plate, the wire also tends to move towards the center of the plate. Subsequent simulation shows that the force in Y-direction goes towards zero when it is exactly opposite of the center of the plate. This is similar to an analytical result on lateral Casimir forces of a sphere and a wire, where the sphere also experiences a lateral force moving it towards the center of the plate[13].

**Fig. 7** shows the Casimir potential for a groove-like structure. Interestingly, computing the forces with the code, the forces on the left object are $f_x=-5 \times 10^{-12}$ $Nm^{-1}$ and $f_y$=0 $Nm^{-1}$ with opposite signs for the right object. Hence, both objects repel each other rather than attracting. Although repulsive Casimir configurations are described in the literature, for instance in the configuration of two halves of thin metal spherical shells (see Ref. 11 for a comprehensive review) or by using strong magnetic polarizability[9], repulsive Casimir forces have not been reported so far for groove-like structures to the knowledge of the author.

## Conclusion

A 3D finite-element code was developed to quickly assess Casimir forces in arbitrary structures. The numerical results were compared to analytical equations showing that they behave similar with respect to distance between the objects. The offset in the absolute value was about one order of magnitude without renormalization, an effect that is well known in the literature.

The code was applied on previously not assessed geometries such as a wire-plate or groove configuration. The wire experienced a lateral force parallel to the plate moving it towards the center of the plate, whereas the groove structure revealed repulsive Casimir forces. These examples demonstrate the usefulness of such a Finite-Element Casimir numerical simulation to quickly assess Casimir forces and their behaviour for arbitrary geometries – and even find new properties related to Casimir forces in previously unaccessed configurations.

## References

[1]Casimir, H.B.G., Proc. Ned. Akad. Wet., **51**, 1948, pp. 793

[2]Casimir, H.B.G., and Polder, "The Influence of Retardation on the London-van der Waals Forces", Physical Review, **73**, 1948, pp. 360-372

[3]Lamoreaux, S.K., "Demonstration of the Casimir Force in the 0.6 to 6 μm Range", Physical Review Letters, **78**(1), 1997, pp. 5-8

[4]Chan, H.B., Aksyuk, V.A., Kleinman, R.N., Bishop, D.J., and Capasso, F., "Quantum Mechanical Actuation of Microelectromechanical Systems by a Casimir Force", Science, **291**, 2001, pp. 1941-1944

[5]Tacsi, E.S., and Erkoc, S., "Simulation of the Casimir-Polder Effect for Various Geometries", International Journal of Modern Physics C, **13**(7), 2002, pp. 979-985

[6]Krech, M., and Landau, D.P., "Casimir Effect in Critical Systems: A Monte-Carlo Simulation", Physical Review E, **53**(5), 1996, pp. 4414-4423

[7]Maclay, J.G., "Analysis of Zero-Point Electromagnetic Energy and Casimir Forces in Conducting Rectangular Cavities", Physical Review A, **61**, 2000, 052110

[8]Rutherford Appelton Laboratory, "Algorithms for Casimir Force Prediction in Complex and Time-Dependent Geometries", ESA Contract Report 15615/01/NL/LvH, Noordwijk, 2003

[9]Kenneth, O., Klich, I., Mann, A., and Revzen, M., "Repulsive Casimir Forces", Physical Review Letters, **89**(3), 2002, 033001

[10]Bezerra, V.B., Klimchitskaya,G.L., and Romero, C., "Surface Roughness Contribution to the Casimir Interaction Between an Isolated Atom and a Cavity Wall", Physical Review A, **61**, 2000, 022115

[11]Bordag, M., Mohideen, U., and Mostepanenko, V.M., "New Developments in the Casimir Effect", Physics Reports, **353**, 2001, pp.1-205

[12]Lamoreaux, S.K., "Calculation of the Casimir force between imperfectly conducting plates", Physical Review A, **59**(5), 1999, pp. R3149-R3153

[13]Klimchitskaya, G. L., Zanette, S. I., and Caride, A. O., "Lateral projection as a possible explanation of the nontrivial boundary dependence of the Casimir force", Physical Review A, **63**(1), 2001, pp. 014101

| | |
|---|---|
| Dimensions of Simulation Domain (X, Y, Z) | 10 μm, 10 μm, 10 μm |
| Number of Gridpoints (X, Y, Z) | 100, 100, 100 |
| Polarizability (Au) | $1.88 \times 10^{-30}$ $m^3$/Atom |
| Density (Au) | $5.9 \times 10^{28}$ Atoms/$m^3$ |

**Table 1** Simulation Parameters

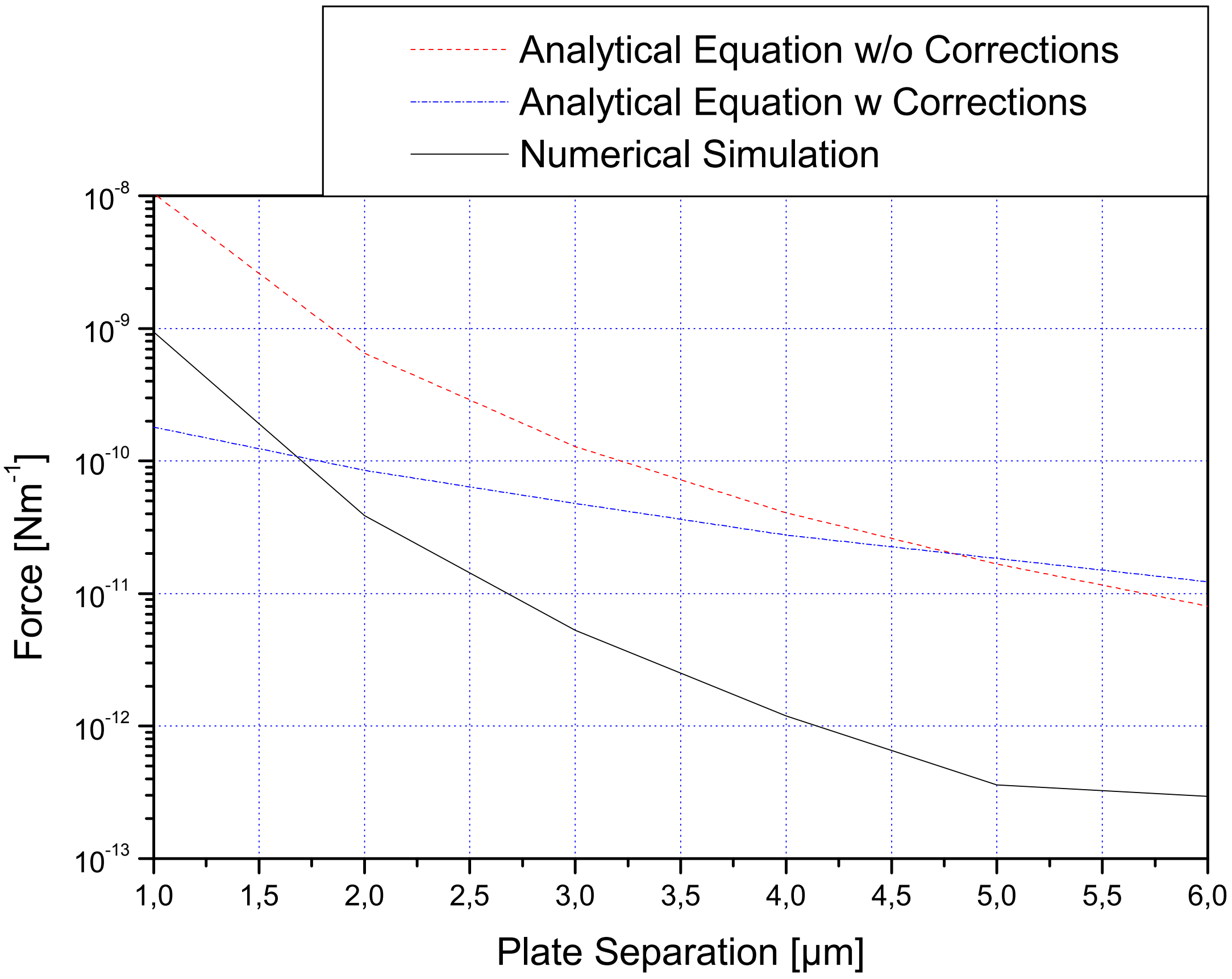

**Fig. 1** Comparison of Numerical Simulation with Equ. (7) and Equ. (11) for a Plate Thickness of 1 µm and a Plate Height of 9 µm

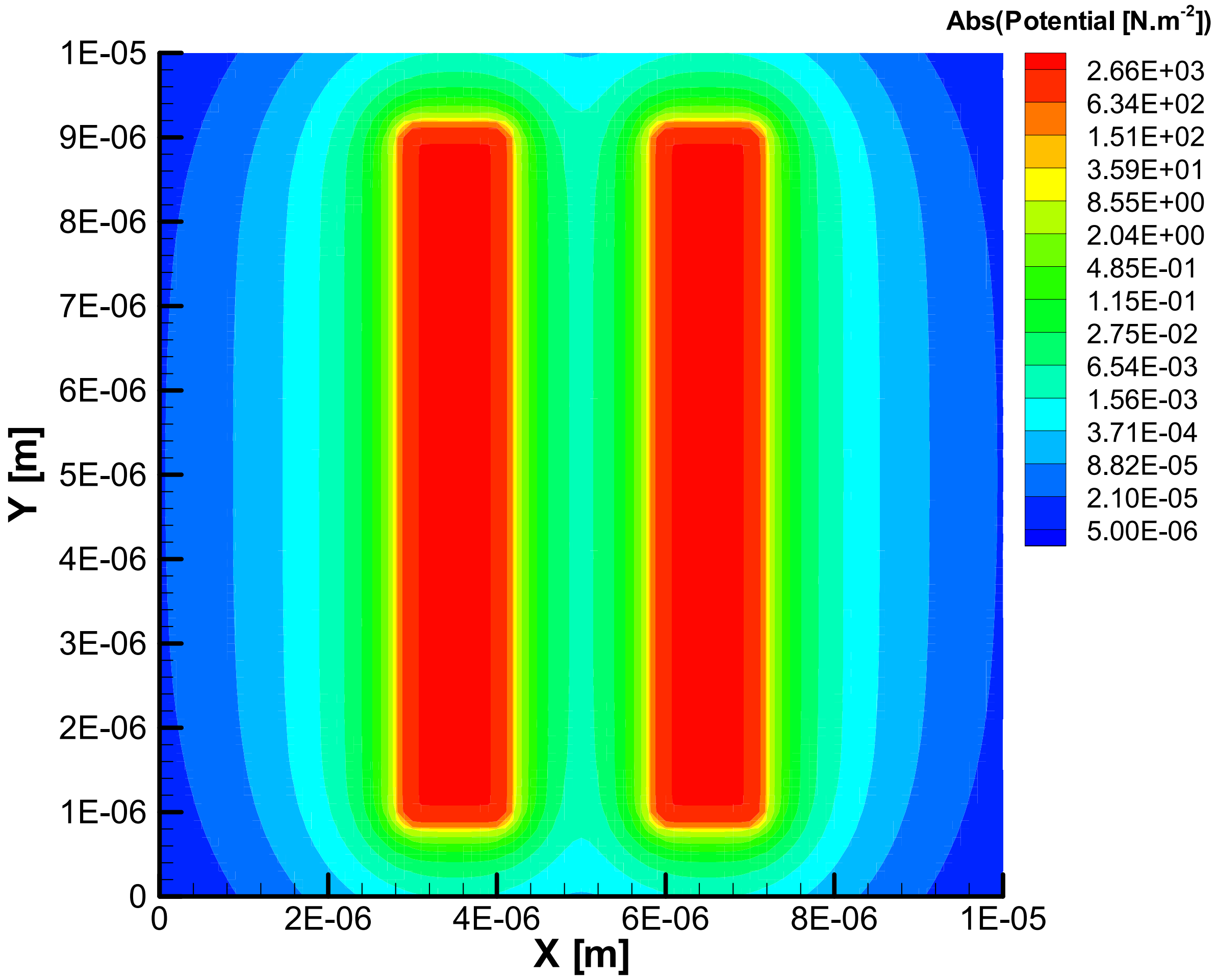

**Fig. 2** Absolute Casimir Potential of Two-Plate Configuration with a Separation Distance of 2 μm in Simulation Domain of 30x30x30 μm

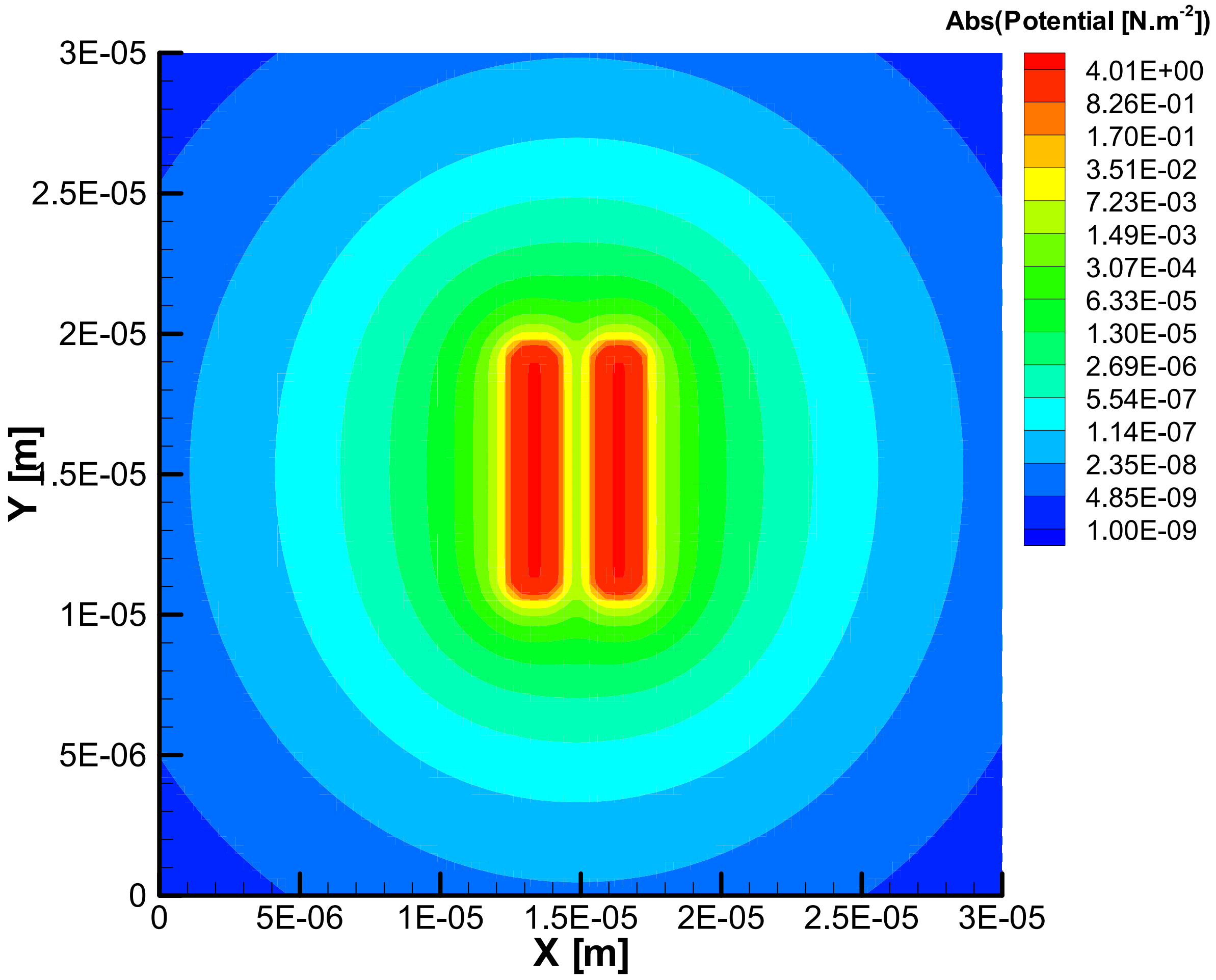

**Fig. 3** Absolute Casimir Potential of Two-Plate Configuration with a Separation Distance of 2 μm in Simulation Domain of 30x30x30 μm

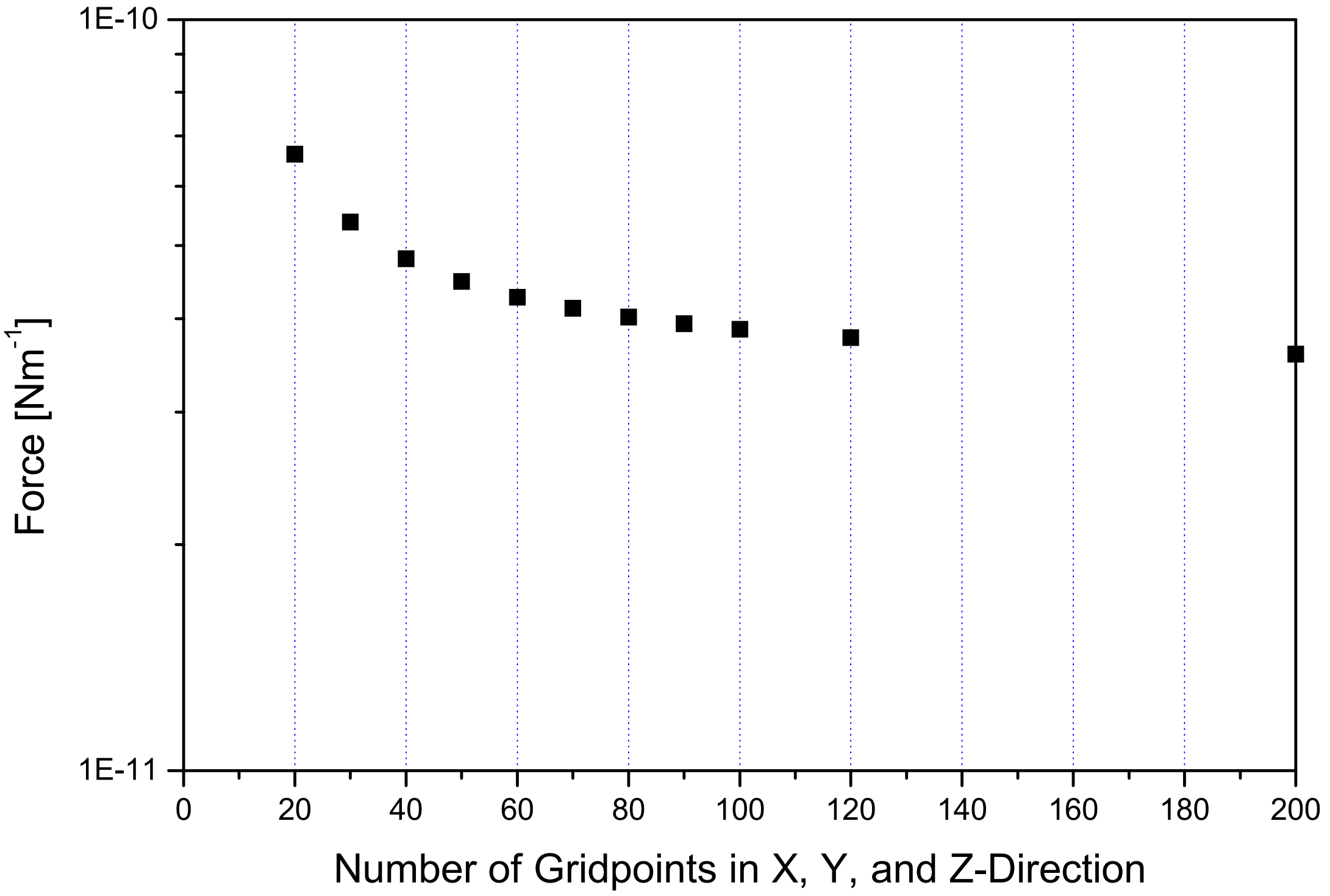

**Fig. 4** Influence on the Casimir Force of Two-Plate Configuration with a Separation Distance of 2 µm on Number of Gridpoints

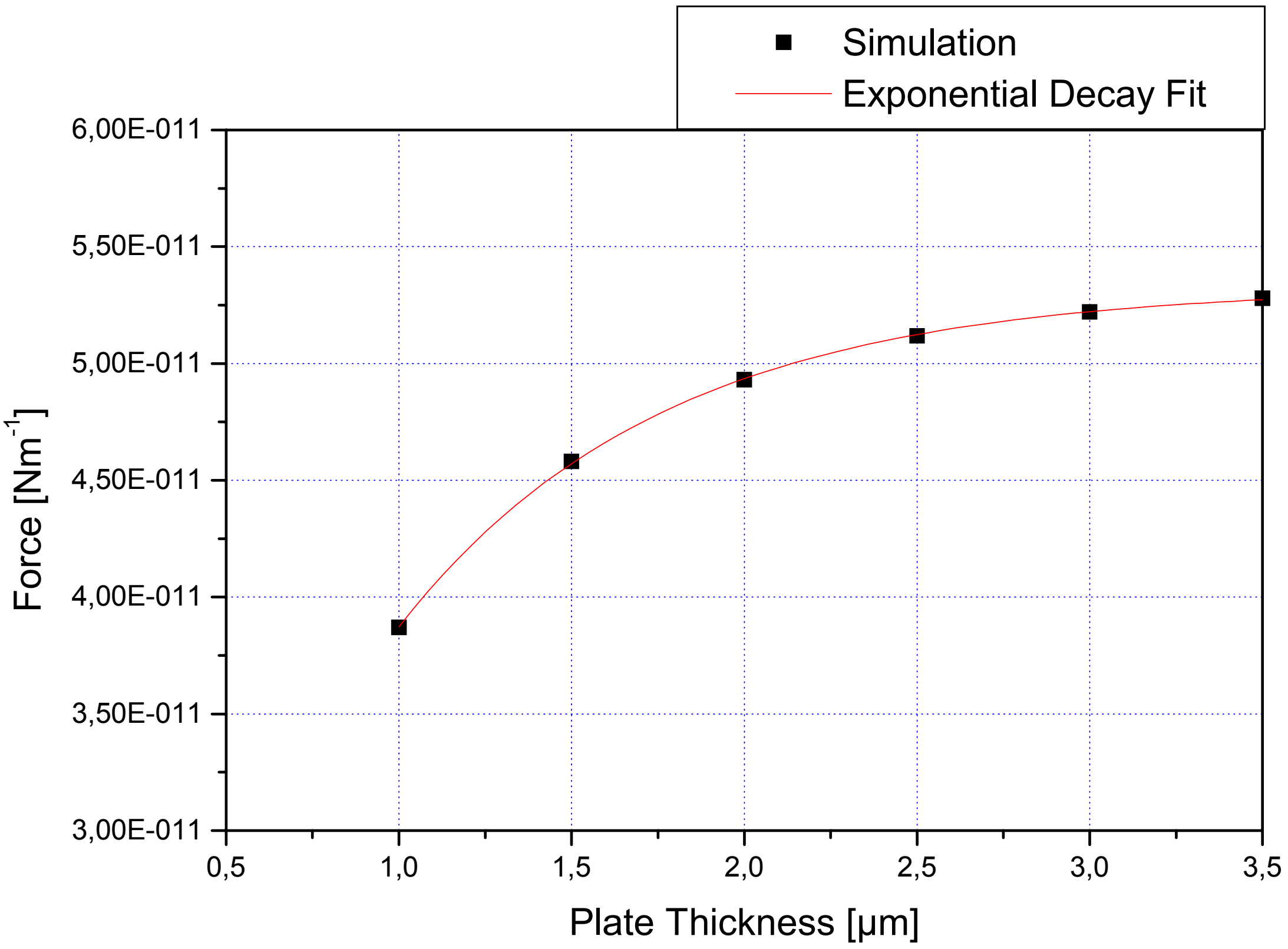

**Fig. 5** Influence on the Casimir Force of Two-Plate Configuration with a Separation Distance of 2 µm on Plate Thickness

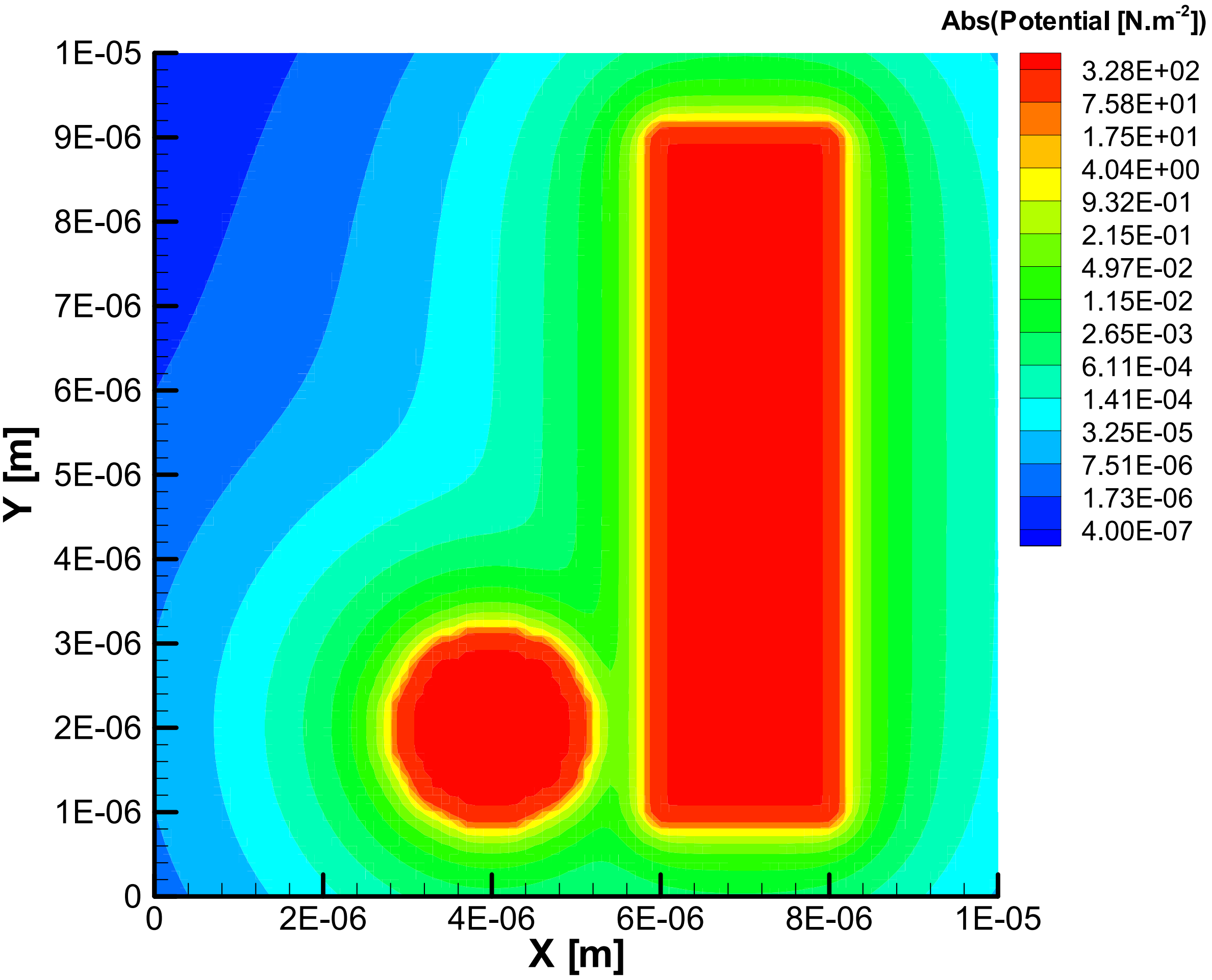

**Fig. 6** Absolute Casimir Potential of Wire-Plate Configuration

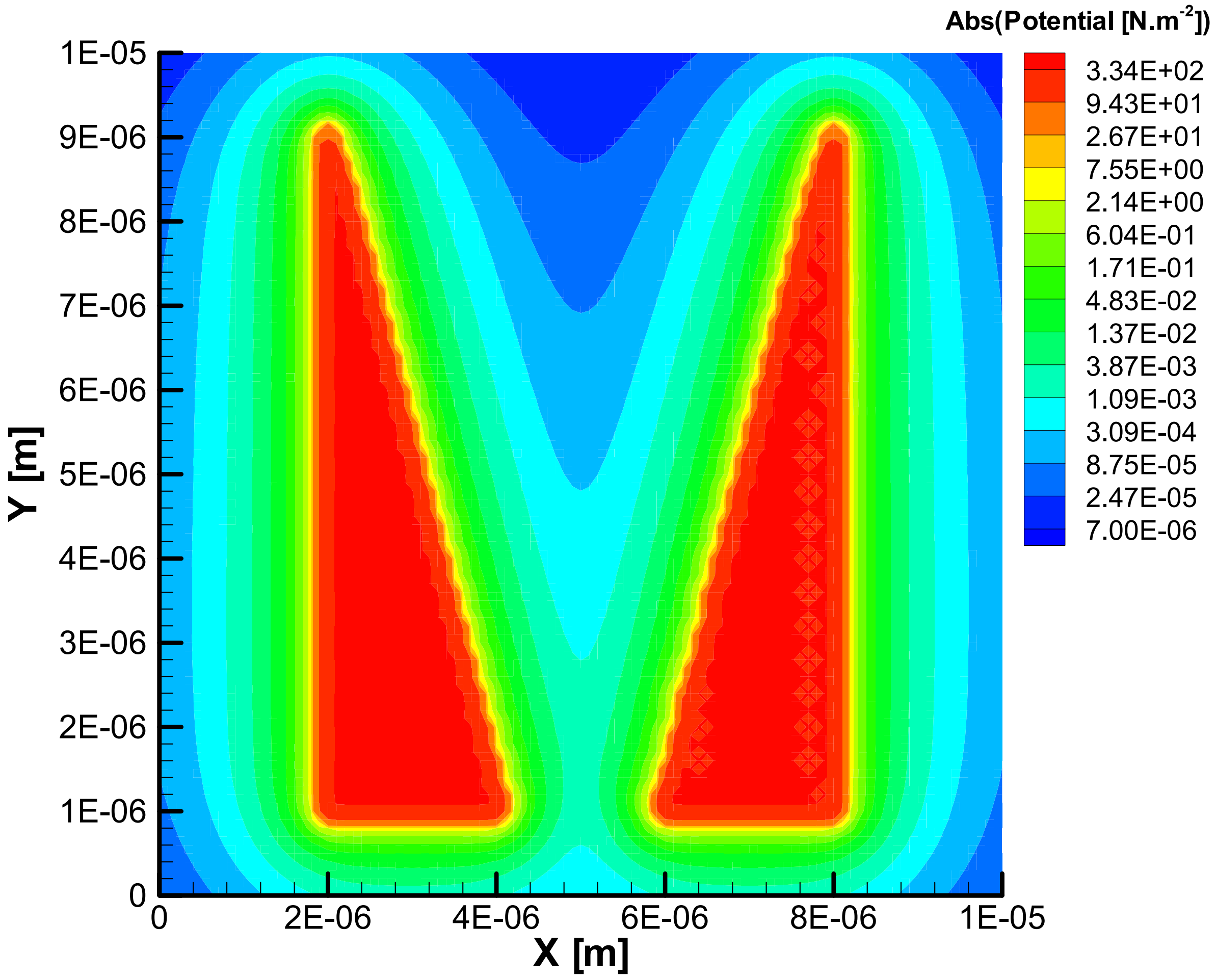

**Fig. 6** Absolute Casimir Potential of Groove Configuration